\documentclass{Interspeech}
\usepackage[bb=boondox,bbscaled=.95,cal=boondoxo]{mathalfa} 

\interspeechcameraready

\title{JIS: A Speech Corpus of Japanese Idol Speakers with Various Speaking Styles}

\author[affiliation={}]{Yuto}{Kondo}
\author[affiliation={}]{Hirokazu}{Kameoka}
\author[affiliation={}]{Kou}{Tanaka}
\author[affiliation={}]{Takuhiro}{Kaneko}

\affiliation[nocounter]{}{NTT Corporation}{Japan}
\email{yuto.kondo@ntt.com, hirokazu.kameoka@ntt.com, kouef.tanaka@ntt.com, takuhiro.kaneko@ntt.com}

\usepackage{comment}

\begin{document}

\maketitle

\begin{abstract}    
    We construct Japanese Idol Speech Corpus (JIS) to advance research in speech generation AI, including text-to-speech synthesis (TTS) and voice conversion (VC). JIS will facilitate more rigorous evaluations of speaker similarity in TTS and VC systems since all speakers in JIS belong to a highly specific category: “young female live idols" in Japan, and each speaker is identified by a stage name, enabling researchers to recruit listeners familiar with these idols for listening experiments. With its unique speaker attributes, JIS will foster compelling research, including generating voices tailored to listener preferences---an area not yet widely studied. JIS will be distributed free of charge to promote research in speech generation AI, with usage restricted to non-commercial, basic research. We describe the construction of JIS, provide an overview of Japanese live idol culture to support effective and ethical use of JIS, and offer a basic analysis to guide application of JIS.
\end{abstract}

\section{Introduction}
\label{sec:intro}
We construct a new speech corpus, Japanese Idol Speech Corpus (JIS), to advance research in speech generation AI, including text-to-speech synthesis (TTS)~\cite{Oord2016wavenet,kim2021conditional,popov2021grad,Casanova2022yourtts} and voice conversion (VC)~\cite{qian2019autovc,popov2022diffusion,kameoka2024voicegrad}. TTS is a technology that generates speech from text, while VC is a technology that transforms a speaker's voice into another person’s voice without changing the spoken content. Recently, neural network-based TTS and VC systems have been actively proposed, achieving very high speech quality. The initial challenge in TTS and VC research is to generate voices of speakers included in training data~\cite{Oord2016wavenet}. More practical TTS and VC systems have emerged, aiming to generate voices not included in the training data~\cite{Casanova2022yourtts,qian2019autovc}. To train and evaluate such systems, it is desirable to use a speech corpus composed of voices from multiple speakers. In the evaluation phase, speech from both training and non-training speakers is used by splitting the multi-speaker corpus. Listeners evaluate the subjective speech quality and the speaker similarity between reference voices included in the corpus and generated voices.

Freely available multi-speaker speech corpora such as CMU ARCTIC~\cite{kominek2004cmu}, LibriSpeech~\cite{panayotov2015librispeech}, and VCTK~\cite{yamagishi2021cstr} for English, and JVS~\cite{takamichi2020jvs} for Japanese, are often utilized. In speech generation AI research using these speech corpora, there may be an implicitly tolerant evaluation of speaker similarity for two reasons;
The first reason is that these speech corpora consist of voices with a wide range of attribute categories in terms of age, gender, occupation, and so on. If the speaker attributes, such as “young female", are perceived to roughly match simply based on the attributes, speaker similarity may be mistakenly judged as high.
The second reason is that the speakers are anonymized, and the listeners are unfamiliar with the speakers, making it less likely for listeners to feel discomfort with the generated voices. As familiar and unfamiliar voices are identified through distinct processes in the brain~\cite{van1987voice}, and people tend to identify which speaker a given voice belongs to more accurately the more familiar they are with the speaker~\cite{schmidt1985identification}, if listeners familiar with the speakers in the corpus were recruited, more rigorous evaluations of speaker similarity in TTS and VC systems could be conducted, focusing on subtle voice characteristics. The tolerant evaluations can become a bottleneck in the development of speech generation AI that achieves speaker similarity performance truly satisfying to users, as the speakers users genuinely wish to generate may include known speakers within a highly specific category.

\begin{table}[t]
  \caption{Comparison between JVS and proposed JIS.}
  \vspace{-0.3cm}
  \label{tab:jisjvs}
  \centering
  \setlength{\tabcolsep}{2pt} 
  \renewcommand{\arraystretch}{0.8} 
  \begin{tabular}{c| c c}
    \toprule
    &JVS&JIS\\
    \midrule
    Language&Japanese&Japanese\\ \hline \begin{tabular}{c}
Recording\\environment\end{tabular}
    &Studio&\begin{tabular}{c}
A: Studio \\ B: Quiet but unspecified \end{tabular}\\\hline
\# of speakers&$100$&\begin{tabular}{c}A: $61$\\B: $108$ \\Total: $169$ \end{tabular}\\\hline
Speaker attribute& \begin{tabular}{c}Professional\\speakers\\(Various genders\\and ages)\end{tabular}&\begin{tabular}{c}Japanese live idols\\(Young female)\end{tabular}\\\hline
\begin{tabular}{c}Total duration\\after VAD\end{tabular}&$30.2$~h&\begin{tabular}{c}A: $12.5$~h\\B: $4.5$~h \\Total: $17.0$~h \end{tabular}
    \\\hline
    Anonymity&Anonymous&\begin{tabular}{c}
Non-anonymous \\(Stage name)\end{tabular}
    \\
    \bottomrule
  \end{tabular}
  \vspace{-0.8cm}
\end{table}
JIS will facilitate more rigorous evaluations of speaker similarity in TTS and VC systems. All speakers in JIS belong to a highly specific category: “young female live idols"~\cite{2021idolsogo}, who are a key part of Japanese pop culture. Moreover, since each JIS speaker is assigned a commonly used stage name, experiment planners may have the possibility to recruit listeners who are familiar with JIS speakers. A comparison between JVS and JIS is shown in Tab.~\ref{tab:jisjvs}. JIS consists of 169 speakers, with a total duration of 17.0 hours after voice activity detection (VAD)~\cite{dinkel2021voice}. JIS speech data are categorized into two types: Speech A, recorded in studios by professional sound directors, and Speech B, recorded in quiet but unspecified rooms. Speech A includes not only 100 sentences with phoneme-balanced texts but also various types of speech commonly used by the idols in their daily activities. Speech B generally covers the same content as Speech A, though it includes only a partial set of the phoneme-balanced sentences. To our knowledge, JIS is the first non-anonymous multi-speaker speech corpus for speech generation AI research. JIS can be regarded as a unique speech corpus from Japan, where the live idol culture is very vibrant.

With its unique speaker attributes, JIS will foster compelling research that is not yet widely studied. For example, live idols are selected from a large pool of applicants based on the expectation that they will entertain fans through various aspects such as performances, interactions, and their unique character traits. As a result, they are likely to possess distinctive and highly appealing voices. This characteristic is expected to advance speech generation AI that focuses on listener-preferred voices. The characteristics of voices that are perceived as attractive vary by listeners~\cite{suda2024finds}, making generating listener-preferred voices a challenging, meaningful task.

JIS will be distributed free of charge to promote research in speech generation AI. To ensure its ethical and effective use, and proper design of listening experiments, understanding live idols and their fan culture is crucial. This paper provides an overview of Japanese live idol culture to assist with JIS usage, submitted before the dataset release to support the collection of additional idol speech data by other research teams.

The contributions of this study are as follows:
\begin{itemize}
\setlength{\itemsep}{-0.0cm}
  \item  We construct JIS, composed of Japanese live idol voices.
  \item We provide an overview of Japanese live idol culture.
  \item We provide results of basic analysis of JIS data.
\end{itemize}

The structure of this paper is as follows: In Sec.~2, an overview of TTS and VC is provided. Section 3 provides an overview of Japanese live idol culture, and discusses JIS construction. Section 4 presents a basic analysis of JIS. Finally, Section 5 summarizes this paper.

\section{Text-to-speech \& voice conversion}
Neural model-based TTS and VC systems have advanced, achieving high speech quality. The initial challenge in TTS and VC research is to generate voices of speakers included in training data~\cite{Oord2016wavenet}. More practical TTS and VC systems have emerged, aiming to generate voices not included in the training data~\cite{Casanova2022yourtts,qian2019autovc}.

Aiming for a more user-friendly speech generation AI, some researchers have proposed new speech generation models based on diverse inference styles, going beyond the direct speaker reference used in conventional TTS and VC. Some models modify voices based on characteristics described in prompts~\cite{hai2024dreamvoice,chen24q_interspeech}, while FleSpeech~\cite{li2025flespeech} combines prompts, reference audio, and images to control a voice attribute. To further promote the development of speech generation AI that users are eager to use, leveraging speech corpora with likable voices, such as JIS, will be effective.
\section{JIS: Japanese Idol Speech Corpus}
\subsection{Japanese live idol}
This section provides an explanation of Japanese live idols~\cite{2021idolsogo} from the late 2010s to the early 2020s to promote the effective and ethical use of JIS, as well as to guide readers who may someday record additional live idol voices.

Live idols are a key part of Japanese pop culture, entertaining fans through live performances that combine singing, dancing, and unique personalities. This paper focuses on female live idols, such as those in the JIS corpus, though both male and female idols exist. Many idols belong to company-managed groups, collaborating with other members to perform. Due to the large number of management companies, some companies cooperate with JIS construction, while others do not. Idol groups, typically with $3$ to $10$ members, perform live more than twice a week. Each group has a unique group name, such as “FRUITS ZIPPER” and “iLiFE!.” Their songs feature divided singing parts, inspiring datasets for singer diarization~\cite{suda2024fruitsmusic}. Idols use stage names, allowing experiment planners to engage fans familiar with JIS speakers while protecting the speakers' privacy.

In the context of numerous rival groups, each group employs various strategies to increase or maintain its fanbase. For instance, they explore innovative performances and more effective promotional techniques while emphasizing appealing qualities such as cuteness, coolness, and fun. As a result, the performances and identities of each group are highly diverse. Furthermore, many management companies frequently organize live events featuring multiple groups by inviting groups from other management companies. This inter-company collaboration facilitates the efficient collection of numerous live idol voices.

The live idol fanbase is diverse, with age distributions and gender ratios varying across idol groups. Fans not only enjoy live performances but also interact with live idols in Instax photo sessions or on social media platforms like $\mathbb{X}$ (formerly Twitter). Fans often become more attached to specific idols than to other members of the same group. This behavior is often described using the slang term \textit{oshi}~\cite{2024oshi}. The term has now become a generalized, multi-meaning slang in Japan, but here it is used in its earliest original sense~\cite{2024oshi}. Each fan often carries a glow stick that have the image color of the object idol of \textit{oshi} during live performances, or take frequent Instax photos with the idol. Every live idol is the object of someone’s \textit{oshi}, which motivates her idol activities. Furthermore, as a criterion for selecting the object of \textit{oshi}, factors such as personality or live performance, which relate to the voice attribute, have a greater influence than appearance~\cite{2024oshi}. These suggest that the perception of which acoustic features make an idol's voice particularly attractive varies between listeners, which implies that the voices of live idols are well-suited for the acoustic task of generating listener-preferred voices.

Many live idols stream YouTube programs to increase their group's visibility, which could lead to their voices being included in large-scale voice corpora like YODAS~\cite{2023yodas} and COCO-NUT~\cite{2023coconut}. However, these voices are often copyrighted, and using them for generative AI research could result in legal issues, depending on future regulations. In contrast, the copyright for the voice data in JIS has been transferred to our company through a contract, with the speakers informed about its intended use. This ensures safer use of JIS.
\subsection{Construction of JIS}
\begin{figure}[t]
  \centering
  \includegraphics[width=0.45\linewidth]{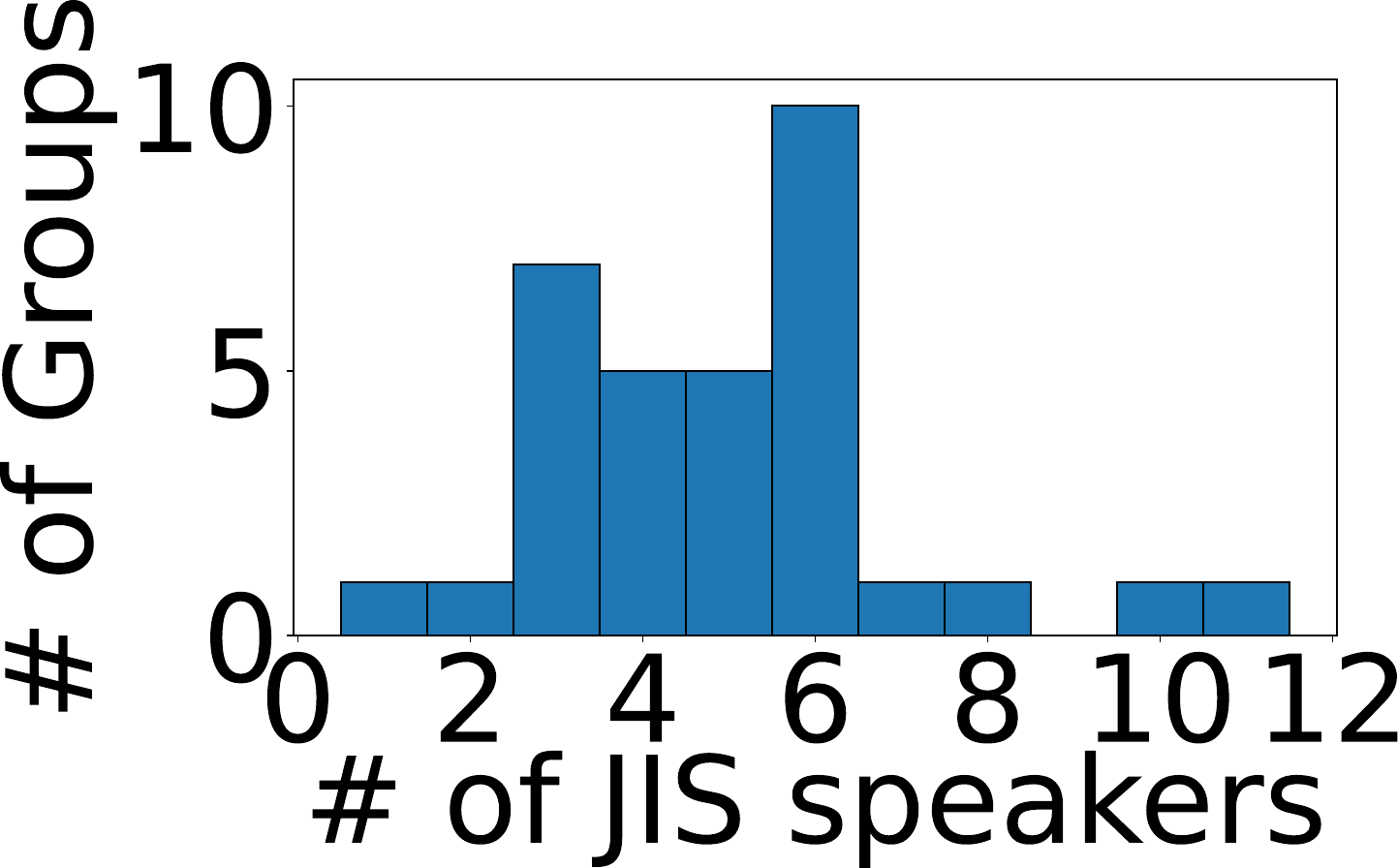}
  \vspace{-0.3cm}
  \caption{Histogram of JIS Speaker Counts by Group.}
  \label{fig:spknum}
  \vspace{-0.7cm}
\end{figure}
The number of JIS speakers was $169$ from $33$ idol groups. The histogram of the number of JIS speakers per group is shown in Fig.~\ref{fig:spknum}. Although the range spans from $1$ to $11$ speakers, most groups have between $3$ and $6$ JIS speakers. $61$ JIS speakers were recorded under the supervision of professional sound engineers in soundproof studios (referred to as Speech A). The remaining $108$ speakers were recorded in quiet rooms selected according to the policies of each group's management company, using the highest quality recording equipment available (referred to as Speech B).
Speech B was recorded with the goal of increasing the number of speakers within a limited budget. All JIS speakers were instructed to produce speech that reflected the character they typically express in their idol activities, while ensuring clear pronunciation as much as possible. All audio files were saved in WAV format, sampled at $48$~kHz, and quantized at $24$~bits.

Both Speech A and Speech B include various types of speech, such as read speech of a phoneme-balanced Japanese corpus, spontaneous speech, greetings for everyday life in Japan, and singing.
As the texts of read speech, we utilized Voice Actress Statistics Corpus (voiceactress100)~\cite{benjo2017voice}, which is identical to “parallel100" in JVS. All speakers read the texts with ruby annotation list~\footnote{\url{https://hiroshiba.github.io/voiceactress100_ruby/}}. For Speech B, only the 10 common sentences from voiceactress100 were utilized.
The spontaneous speech recordings include Japanese phrases such as “Thank you, that was [Group Name]," simulating a farewell greeting after a performance, “Thank you for coming," intended for first-time attendees of an Instax photo session, and a Japanese phrase expressing the speaker's or group's individuality, primarily self-introductions.
The farewell greeting after a performance is usually delivered to large audiences and is expected to be more energetic. The Instax photo session is an important opportunity to connect with fans~\cite{2021idolsogo}, creating a more intimate tone. The phrases expressing individuality or group identity were recorded to capture a variety of expressions.
The phrases of the spontaneous speech are commonly used in daily activities, offering a more natural representation of each speaker's character.
The Japanese greeting recordings include nine phrases, such as “Hello” and “I’m off,” all compiled into a single audio file, separated by silent intervals.
The singing recordings include the Japanese folk song {\it Katatsumuri}, a royalty-free track that used in JVS-MuSiC dataset~\cite{tamaru2020jvs}.

Supplementary information was collected for all JIS speakers, except for the three idols in Speech B. This includes the idol group's name, the stage name of the idol, information on prefecture of origin, and questionnaire results for each speaker. In the questionnaire, each speaker was asked to describe their impression of the voices of all speakers belonging to the same group (including their own).
\subsection{Copyright of data and limitations for use }
The copyright for the collected speech data has been transferred to our company under a contract that limits the scope of its use. The usage of JIS is restricted to non-commercial, basic research.

JIS will be distributed free of charge to other research institutions that agree to use JIS in accordance with our company's terms and for ethical academic research purposes. Users must be mindful that the idols are actively involved in the entertainment industry, and care must be taken to avoid harming the reputation of the recorded groups, their management companies, fans, or any associated parties. Note that the appealing image of live idols must be maintained so as not to disappoint fans~\cite{galbraith2012introduction}.

\section{Analysis of JIS}
This section presents the analysis of JIS. In Sec.~\ref{sec:analothers}, analysis of supplementary information is discussed, and in Sec.~\ref{sec:analspeech}, analysis of the speech data is conducted.
\subsection{Analysis of supplementary information}
\label{sec:analothers}
The analysis results of supplementary information will provide useful guidelines for designing listening experiments and conditioning speech processing models using JIS.
\subsubsection{Prefecture of origin}
\begin{table*}[t]
  \caption{JIS Speaker Counts by Region of Origin in Japan}
  \vspace{-0.3cm}
  \label{tab:origin}
  \centering
  \renewcommand{\arraystretch}{0.3} 
  \begin{tabular}{c c c c c c c c c c}
    \toprule
    Hokkaido & Tohoku & Kanto & Chubu & Kansai & Chugoku & Shikoku & Kyushu\&Okinawa & Foreign countries&Hidden \\
    \midrule
    $5$ & $12$ & $86$ & $28$ & $15$ & $6$ & $2$ & $8$ & $2$&$2$ \\
    \bottomrule
  \end{tabular}
  \vspace{-0.5cm}
\end{table*}

The information about JIS speakers' regions of origin serves as both a reference for regional bias and a useful auxiliary input for speech processing models, such as in Japanese dialect-based TTS systems~\cite{Yamauchi2024hogentts}.
Based on the obtained prefecture information, the number of JIS speakers in each regional division was aggregated. The standard eight regional divisions~\cite{regionJapan} were used. Note that these divisions do not necessarily correspond to dialect regions. Various dialectical region divisions based on indicators such as phonology, accent, and vocabulary exist~\cite{Abe2015hogen}.

Table \ref{tab:origin} shows the number of JIS speakers from each region. About half of the JIS speakers are from areas outside of the Kanto region including Tokyo, confirming that speakers from various regions are included. JIS speakers from the United States and Taiwan were also identified, with one speaker from each region.
\subsubsection{Questionnaire result on voice impression}
\begin{figure}[t]
  \centering
  \includegraphics[width=1.0\linewidth]{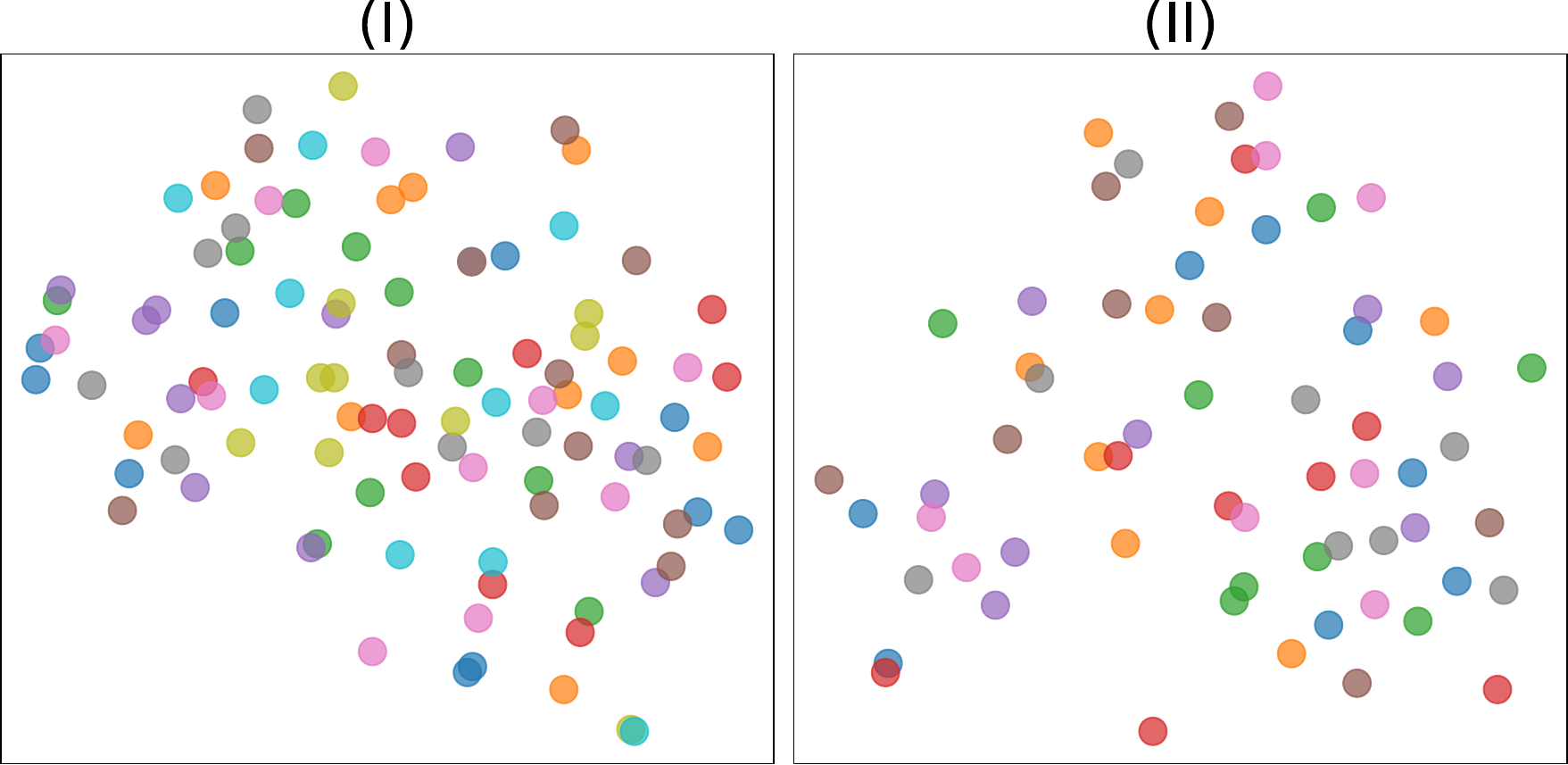}
  \vspace{-0.3cm}
  \caption{Embedding plots of free descriptions given for JIS speakers of two idol groups ({\rm (I)} and {\rm (II)}).  The same color represents the free descriptions given for the same JIS speaker, but the corresponding speakers differ between {\rm (I)} and {\rm (II)}.}
  \vspace{-0.4cm}
  \label{fig:bert}
\end{figure}
The Japanese free-text descriptions of the voice impressions were morphologically analyzed. Japanese adjectives corresponding to expressions such as “energetic," “calm," “bright," “kind," and “transparent" frequently appeared. Using the Japanese BERT model~\cite{huggingface_bert_japanese}, $768$-dimensional latent variables were extracted from all the evaluations of two idol groups and visualized using t-SNE. The results are shown in Fig.~\ref{fig:bert}. Figures \ref{fig:bert} {\rm (I)} and {\rm (II)} correspond to the $100$ and $64$ free-text evaluations, respectively, from groups with $10$ and $8$ JIS speakers shown in Fig.~\ref{fig:spknum}.
Note that the same color represents the free descriptions given for the same JIS speaker, but the corresponding speakers differ between {\rm (I)} and {\rm (II)}. In {\rm (I)} and {\rm (II)}, clouds of same color points are mixed with those of other color points. This indicates that for the same JIS speaker, a variety of free-text impressions were obtained. Moreover, the overlapping yet offset regions of the point clouds suggest differences in voice impressions between JIS speakers.
\subsection{Analysis of speech data}
In this section, we analyze JIS speech data. The analysis results will provide valuable guidelines for use of JIS speech data.
\label{sec:analspeech}
\subsubsection{Pseudo MOS}
\begin{figure}[t]
  \centering
  \includegraphics[width=0.8\linewidth]{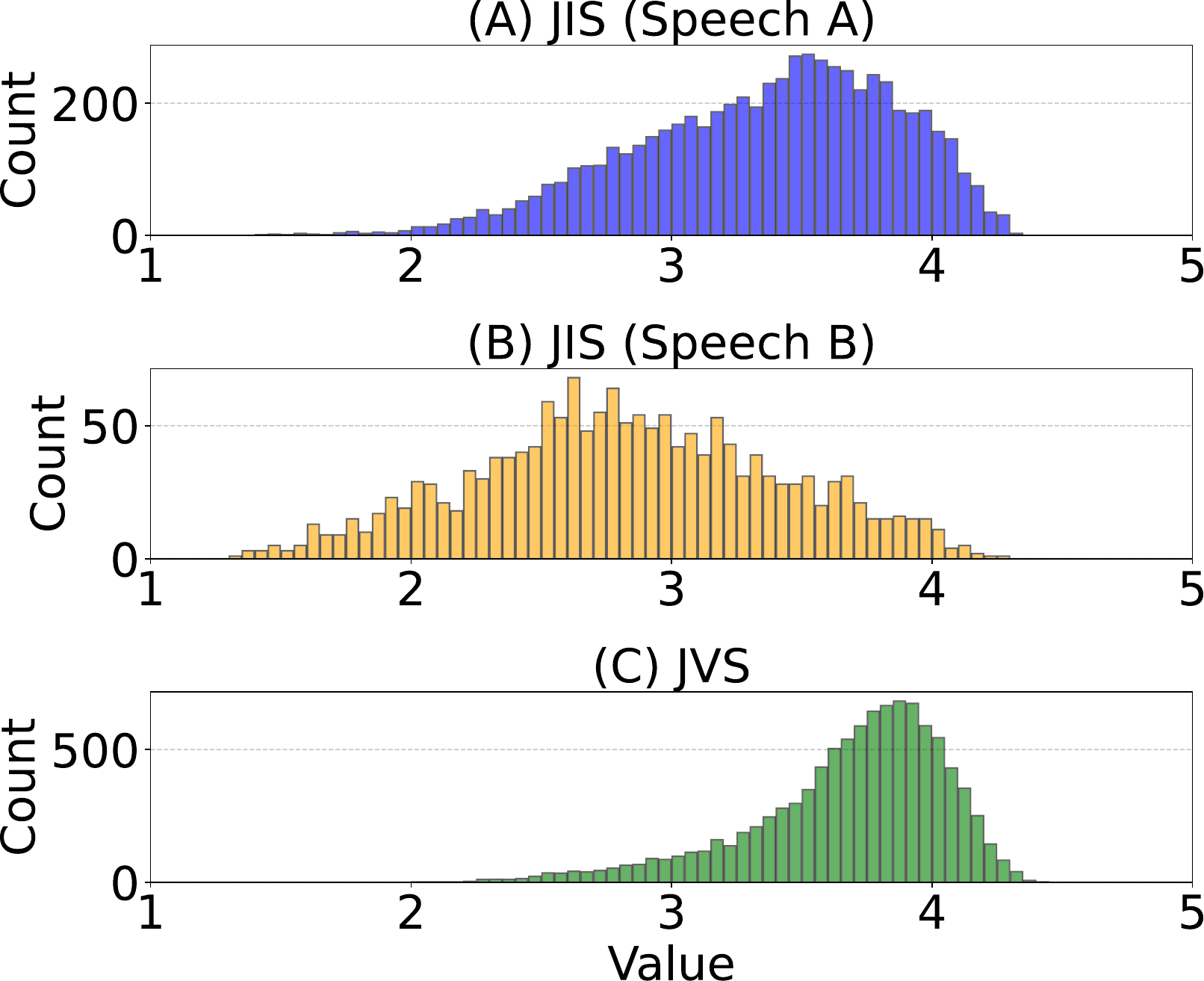}
  \vspace{-0.3cm}
  \caption{Histograms of pseudo MOS.}
  \vspace{-0.4cm}
  \label{fig:pmos}
\end{figure}
To roughly evaluate the decline in subjective speech quality of Speech B, which were recorded in quiet but unspecified rooms, compared to Speech A, which were recorded in studios, we calculated the pseudo mean opinion score (MOS) using the officially available UTMOS~\cite{saeki2022utmos} model. UTMOS is a model that predicts MOS of speech generated by TTS or VC systems in terms of subjective speech quality and has achieved strong results in VoiceMOS Challenge 2022~\cite{huang2022voicemos}. We preprocessed JIS audios by normalizing based on the average volume after VAD and downsampling to $16$~kHz. We tested various normalization constants, choosing the one with the highest average pseudo MOS.

The histograms of all the pseudo MOS values for Speech A, Speech B, and “parallel100" subset of JVS for comparison are shown in Fig.~\ref{fig:pmos}. The mean pseudo MOS values for Speech A, Speech B, and JVS were $3.4$, $2.8$, and $3.7$, respectively. The standard deviations of the pseudo MOS values for Speech A, Speech B, and JVS were $0.50$, $0.58$, and $0.37$, respectively. The average values of Speech A and JVS, which are recorded in studios, are somewhat low, likely due to the linguistic mismatch with the speakers used in the UTMOS training data. The lower average and larger standard deviation for Speech A compared to the JVS average may result from speech hesitations, as JIS speakers are not professional speakers. Since a large portion of the histograms overlap between Speech A and Speech B, Speech B may contain studio-quality audio samples. The large standard deviation for Speech B is likely due to the more diverse recording environments in which Speech B were collected.
\subsubsection{Speaker embedding}
\begin{figure}[t]
  \centering
  \includegraphics[width=1.0\linewidth]{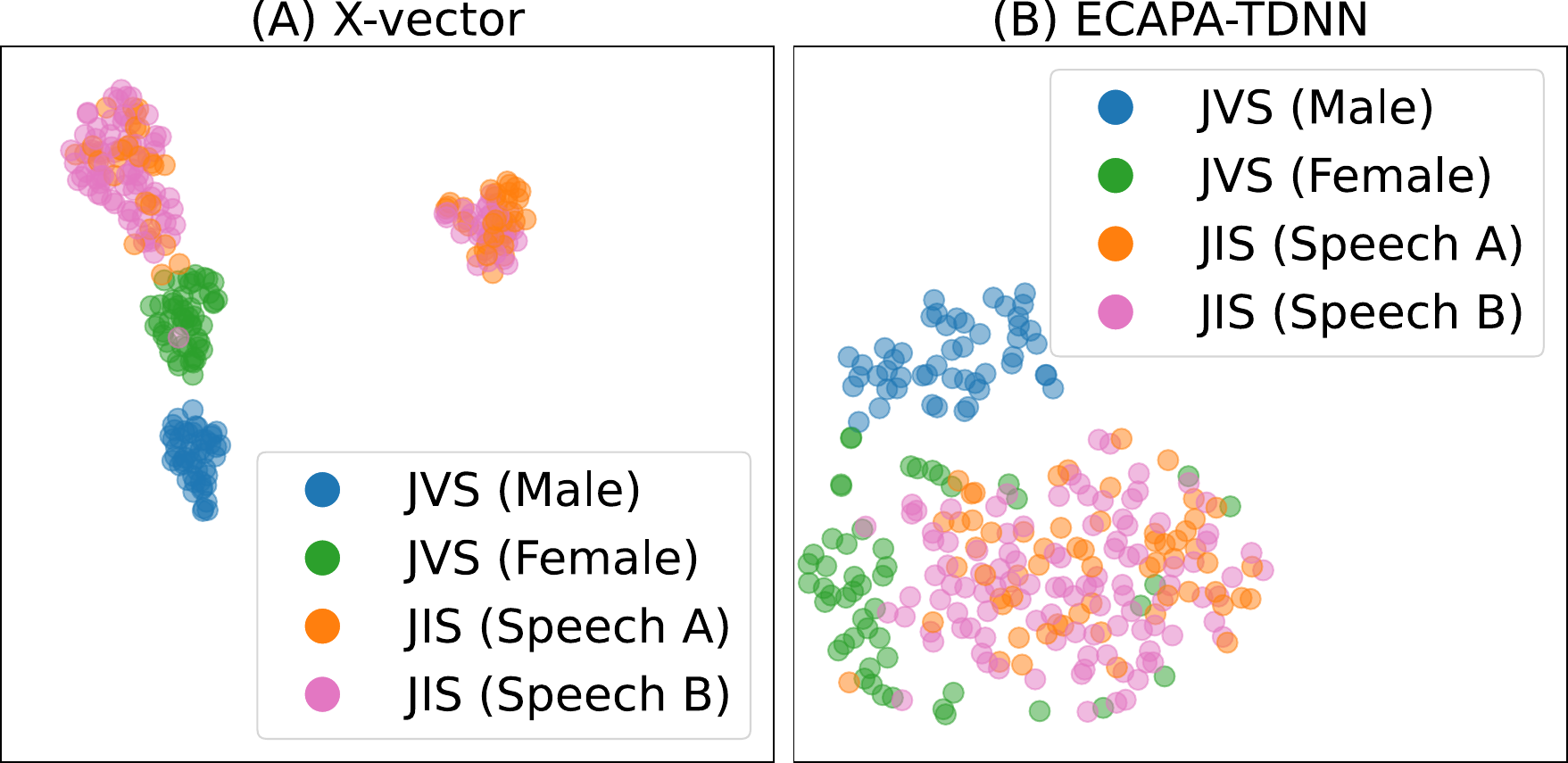}
  \vspace{-0.6cm}
  \caption{Embedding plots of JIS and JVS speech data.}
  \label{fig:jisjvsplot}
  \vspace{-0.45cm}
\end{figure}
\begin{figure}[t]
  \centering
  \includegraphics[width=0.9\linewidth]{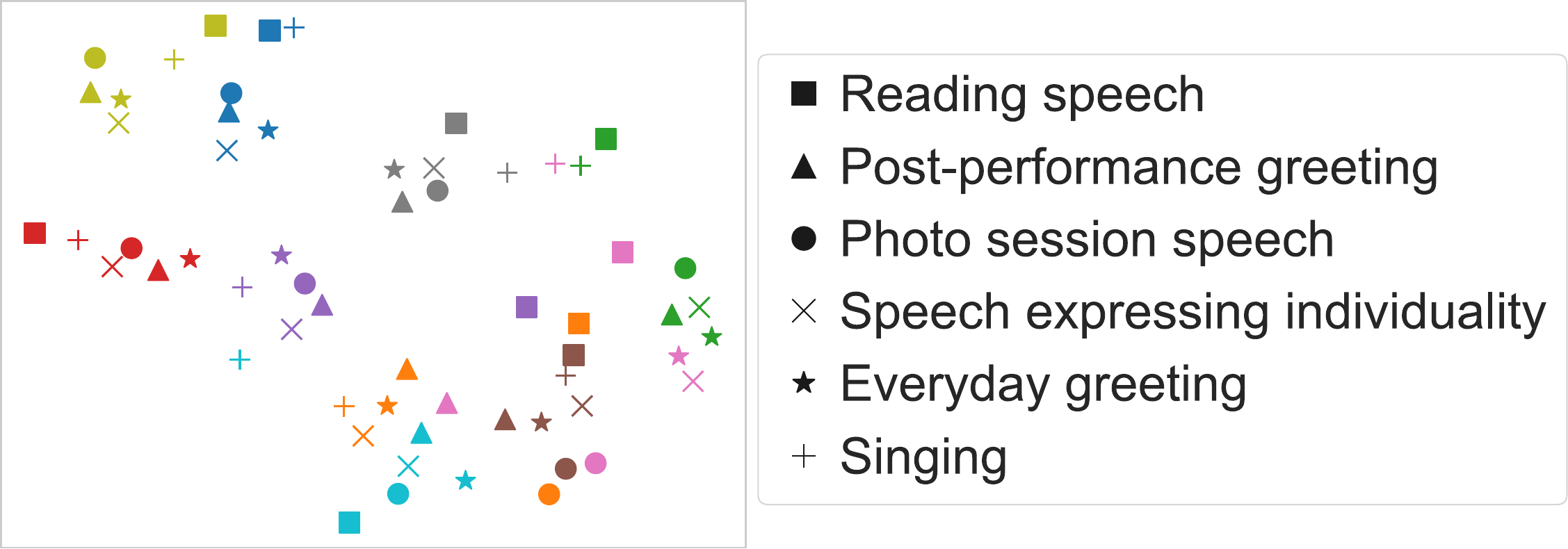}
  \vspace{-0.3cm}
  \caption{Embedding plot of various speaking styles of JIS speakers in one idol group. The same color represents the same JIS speaker.}
  \label{fig:10jisplot}
  \vspace{-0.8cm}
\end{figure}
To examine the differences in speaker individuality between JVS and JIS speakers, we visualized speaker embeddings. We used x-vector~\cite{2018xvector} and ECAPA-TDNN~\cite{desplanques2020ecapa} trained on Japanese speech data~\cite{xvectorjap,ecapajap}. These models were trained on $1233$ and $4960$ speakers, respectively.

All speaker embeddings of the sentence number “001" of voiceactress100 after t-SNE are shown in Fig.~\ref{fig:jisjvsplot}. Both x-vector and ECAPA-TDNN embeddings reveal distinct clusters for JVS male speakers, JVS female speakers, and JIS speakers, which reveal that JIS speakers have different attributes compared to JVS speakers. The JIS cluster is closer to the JVS female cluster and farther from the JVS male cluster, reflecting that all JIS speakers are young females. Both plots show that Speech A and Speech B are mixed together, indicating that the differences in recording environments have not significantly affected the output of these models.

Figure \ref{fig:10jisplot} illustrates ECAPA-TDNN embeddings across various speaking styles after t-SNE, using the idol group including $10$ JIS speakers. The points tend to form clusters for each speaker, suggesting that the instruction to produce speech reflecting their typical idol persona was somewhat effective in leading to consistent speaker-specific speech. For some speakers, the points corresponding to reading speech are located outside their respective speaker clusters, likely due to difficulty in reading phoneme-balanced sentences. In several speaking styles, points for some speakers are densely concentrated. This may be attributed to the characteristic tendencies of speech impressions associated with each speaking style, such as the energetic nature of post-performance greeting and the intimate tone of photo session speech. Such concentration is not observed in speech expressing individuality or in every greeting, suggesting that the impressions of these speech are more diverse.
\vspace{-0.5cm}
\section{Conclusion}
We constructed a new voice corpus, JIS, composed of Japanese live idol voices. JIS will be distributed free of charge to other research institutions. To encourage the effective and ethical use of JIS and provide guidance for future efforts to collect live idol voices, we provided an overview of Japanese live idol culture. We analyzed JIS to support JIS usage.

Future work includes conducting questionnaire on JIS voices as preparation for generating listener-preferred voices, and constructing a speech corpus composed of speakers in another highly specific category, such as “young male live idols."
\section{Acknowledgement}
This work was supported by JST CREST Grant Number JP-MJCR19A3.
\bibliographystyle{IEEEtran}
\bibliography{kondobib}

\end{document}